# Precision measurement of the ground-state hyperfine constant for $^9Be^+$ in a linear Paul trap via magnetically insensitive hyperfine transitions

Zhi-yuan Ao [1,2,*] Wen-li Bai [1,2,*] Qian-yu Zhang [3] Wen-cui Peng [1,†] and Xin Tong [1,‡]
[1] Innovation Academy for Precision Measurement Science and Technology, Chinese Academy of Sciences, Wuhan 430071, People's Republic of China
[2] University of the Chinese Academy of Sciences, Beijing 100049, People's Republic of China
[3] School of Electronics and IoT, Chongqing Polytechnic University of Electronic Technology, Chongqing 401331, People's Republic of China

(Received 22 January 2026; accepted 10 April 2026; published 27 April 2026; corrected 4 August 2026)

Direct measurements of the ground-state magnetically insensitive hyperfine transition $|F=2, m_F=0\rangle \rightarrow |F=1, m_F=0\rangle$ of $^9Be^+$ ions have been performed using microwave-driven state transfer. The $^9Be^+$ ions are confined and laser cooled in a linear Paul trap, forming a Coulomb crystal. The transition frequencies have been measured over a magnetic-field range of $\pm 0.5$ mT centered at zero magnetic field and the acquired data were fitted accounting for the high-order Zeeman effect. The hyperfine constant $A$ is determined to be $-625.008\,840(35)$ MHz, achieving a relative precision of $5.6\times10^{-8}$.

## I. INTRODUCTION

The beryllium ion ($^9Be^+$) is an important atomic system with a unique electronic structure and energy-level properties, making it indispensable in precision measurements [1,2], sympathetic cooling experiments [3,4], ultracold reaction dynamics [5], and quantum information science [6,7]. As the lightest alkali-metal-like ion with three electrons, its energy levels can be calculated with high theoretical accuracy [8,9], enabling stringent tests of quantum electrodynamics (QED) through comparisons of experimental results and theoretical predictions. For the $^9Be^+$ ion, its ground state serves as a sensitive probe of nuclear structure parameters, including the nuclear magnetic moment, charge distribution, and magnetization distribution. Consequently, precise measurements of its ground-state hyperfine structure provide sensitive tests of nuclear theory models [1,10,11].

The ground-state hyperfine splitting arises from the interaction between the nuclear magnetic dipole moment and the magnetic field generated by the valence electron at the nucleus. This interaction is quantified by the hyperfine constant $A$. Precision measurements of the ground-state hyperfine splitting have been performed on multiple species, such as $Be^+$ [12,13], $Cd^+$ [14], $Mg^+$ [15], $Ba^+$ [16,17], $Yb^+$ [18–21], and $Hg^+$ [22]. The most precise experimental determinations of the $A$ value to date have been performed at tesla-level magnetic fields in Penning traps, reaching an accuracy on the order of $10^{-11}$ [12,13]. At this level of precision and field strength, measurements have resolved the second-order Zeeman effect correction term—the diamagnetic correction, which scales quadratically with $B^2$ [13,23]. To accurately model these high-field measurements and extract the zero-field hyperfine constant, a fully relativistic theory describing the Zeeman splitting of the ground-state hyperfine structure levels in lithium-like ions is implemented [24]. The dominant correction terms that deviate from the Breit-Rabi model [25] scale linearly or quadratically with the magnetic-field strength $B$ ($\propto B$ or $B^2$). Therefore, the accurate determination of the external field strength and relative corrections becomes particularly crucial for extracting the value of $A$. It is also worth mentioning that a $2.4\sigma$ discrepancy persists between the most precise $A$ values obtained in tesla-level magnetic-field experiments [13,26].

For measurements performed under weak magnetic fields, an experiment conducted in a linear Paul trap at 0.7 mT measured two magnetically sensitive ground-state hyperfine transitions, yielding a value of $A$ with a relative precision of $4\times10^{-7}$ [27]. In weak magnetic fields, $B$-dependent corrections are effectively suppressed. Nevertheless, the achievable precision under weak magnetic fields remains inherently limited by the magnetic sensitivity of these transitions.

In this work, to suppress the magnetic sensitivity and accurately determine $A$, we report high-precision measurements of the magnetically insensitive transition $^2S_{1/2}$ $|F=2, m_F=0\rangle \rightarrow |F=1, m_F=0\rangle$ for $^9Be^+$ Coulomb crystals confined in a linear Paul trap. Using state preparation with polarized cooling lasers and state transfer via microwave pulses, the resonance frequencies of this transition are measured over a magnetic-field range of $\pm 0.5$ mT centered at zero magnetic field and fitted to a theoretical model incorporating

*These authors contributed equally to this work.
†Contact author: wencuipeng@wipm.ac.cn
‡Contact author: tongxin@wipm.ac.cn

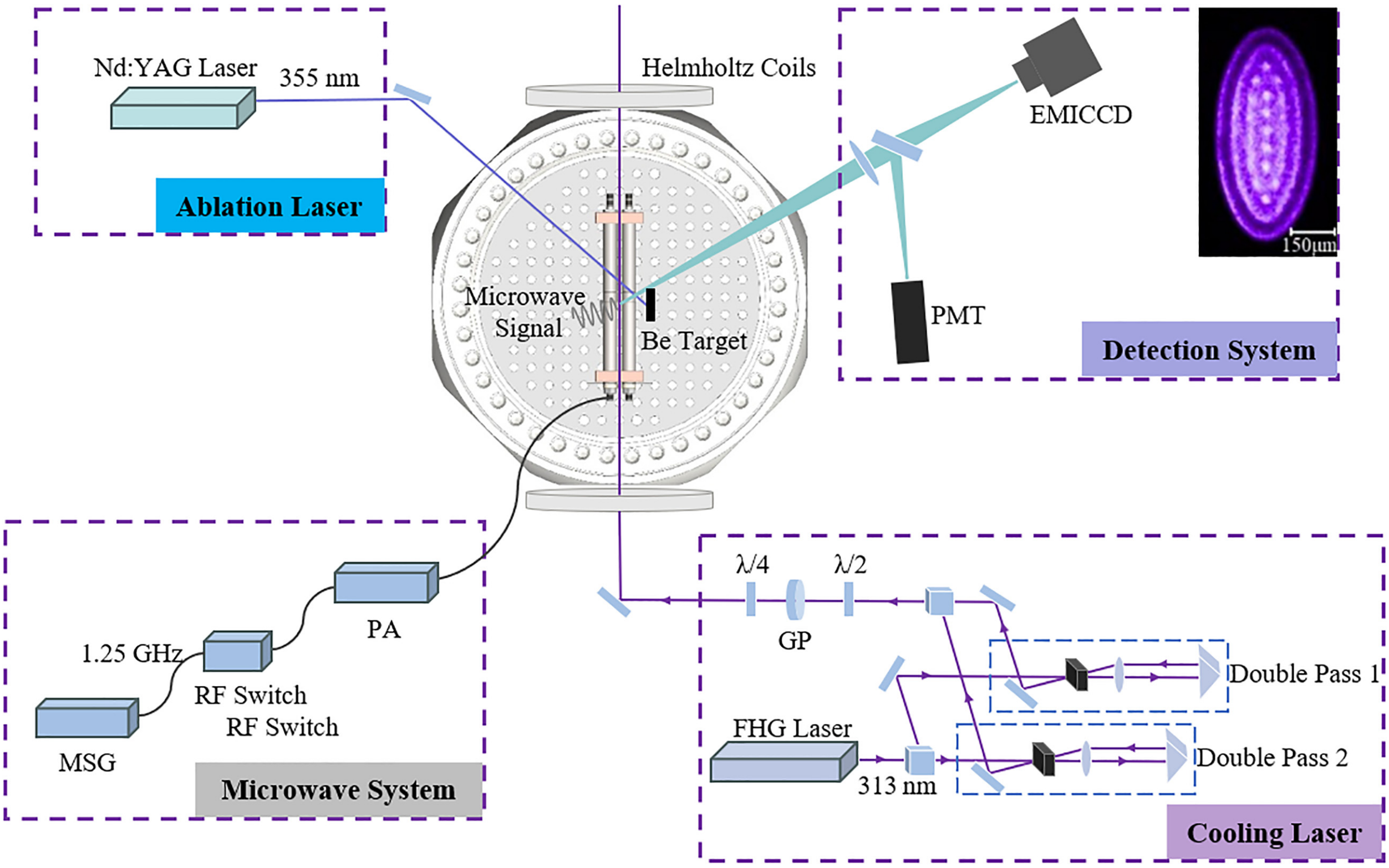

FIG. 1. Experimental setup for microwave spectroscopy of $^9Be^+$ ions. The upper-right inset shows a $^9Be^+$ Coulomb crystal imaged by the EMICCD camera. MSG: microwave signal generator; PA: power amplifier; GP: Glan-Taylor prism; PMT: photomultiplier tube detector; EMICCD: electron-multiplying intensified charge-coupled device.

higher-order Zeeman effects, enabling precise extraction of the hyperfine constant $A$.

## II. EXPERIMENTAL SETUP

The experimental apparatus, schematically illustrated in Fig. 1, consists of five components: an ion trap system, a laser system for ionization and cooling, a fluorescence detection system for transition signal acquisition, a magnetic-field control system, and a microwave system.

The ion trap system consists of a linear Paul trap and a beryllium target positioned adjacent to the trap. The linear Paul trap is constructed with four parallel rods made of 316L stainless steel, with a length of 117.7 mm, a diameter of 8 mm, and a diagonal spacing of 7 mm. Each rod is fitted with endcap electrodes of 24 mm in length [28]. A radio-frequency (RF) voltage with an amplitude of 350 V and a frequency of 13.8 MHz is applied to one pair of diagonal rods, while the opposite pair is grounded. All twelve electrodes can be individually biased with direct-current (DC) voltages to compensate for micromotion and precisely control the geometry and position of Coulomb crystals [28].

The laser system includes an ion ablation unit and a cooling unit. For laser ablation, a commercial pulsed Nd:YAG laser (Beamtech, Nimma-Extra) operating at 355 nm (third harmonic generation) is used to sputter and ionize beryllium atoms from the target [29]. For cooling, a continuous-wave 313 nm laser beam generated by fourth-harmonic generation of a commercial diode laser system (Toptica TA-FHG Pro) serves as the cooling laser. This ultraviolet (UV) beam is split into two paths, each passing through a double-pass acousto-optic modulator (AOM). The AOMs produce two laser beams with a frequency separation of 1.25 GHz, which act as the cooling and repumping beams, respectively. The two beams are recombined and then passed through a Glan-Taylor prism, followed by a quarter-wave plate, and converted into high-purity $\sigma^+$-polarized light before being directed into the vacuum chamber.

In the fluorescence detection system, an EMICCD camera [Princeton Instruments, Model MAX 4–512EM(B)] and a photomultiplier tube (PMT, Thorlabs, Model PMTSS) are employed simultaneously to collect fluorescence signals from Coulomb crystals. The camera images the Coulomb crystal to monitor its shape, size, and structural order. The PMT provides high-sensitivity, time-resolved measurement of fluorescence intensity, which is crucial for state-selective detection.

The magnetic-field control system comprises three pairs of Helmholtz coils mounted outside the vacuum chamber. The axial pair of Helmholtz coils, aligned with the laser propagation direction, is powered by a precision power supply (ITECH IT6412) and generates a tunable magnetic field of up to 5 mT at the ion trap center. The two radial coil pairs, powered separately by two precision power supplies (ITECH IT6723), compensate for the Earth's magnetic field and ambient magnetic field.

Microwave radiation used to drive the hyperfine transitions is generated by a microwave synthesizer (Rohde & Schwarz

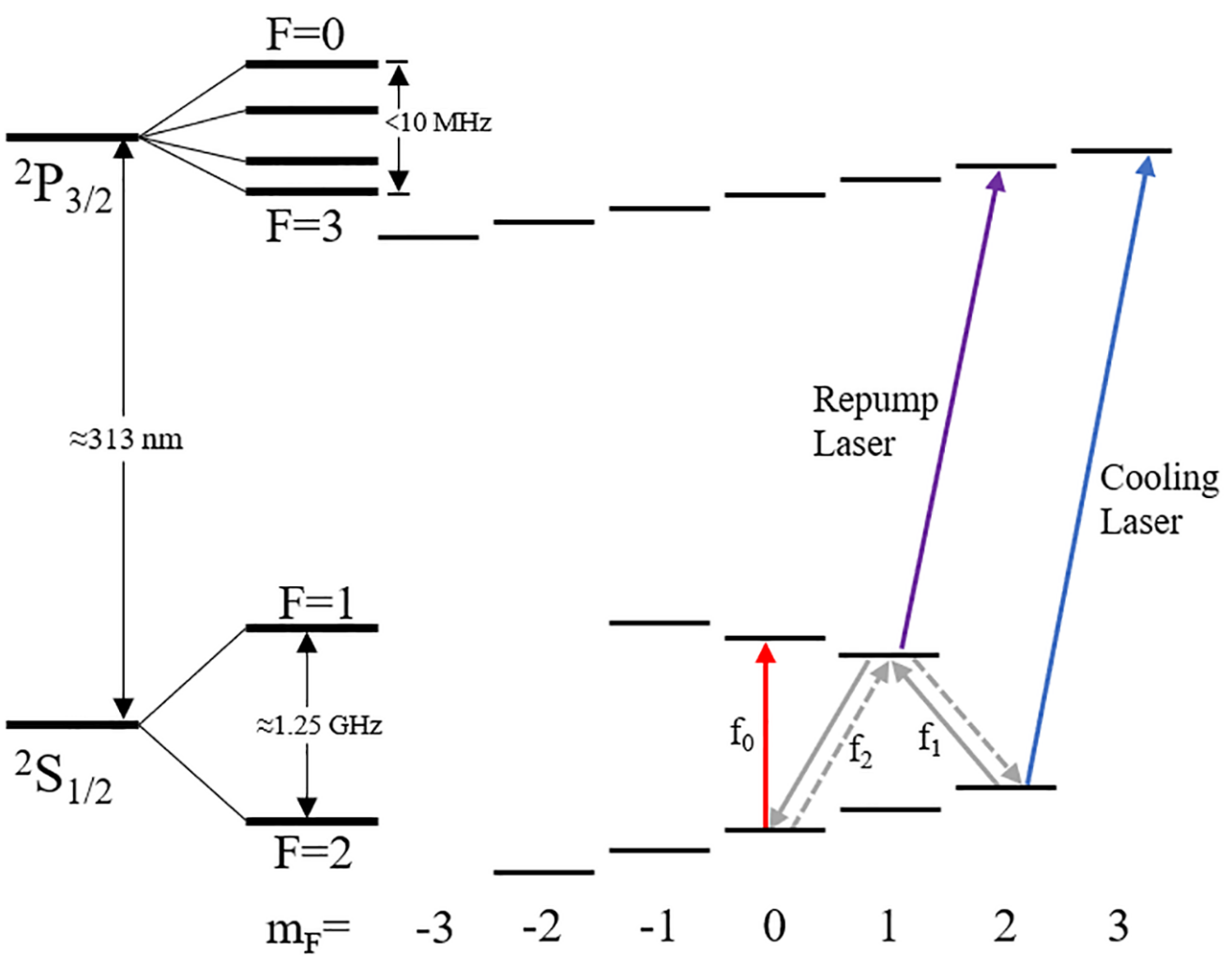

FIG. 2. Energy-level structure of the $^9Be^+$ ion (not to scale). The diagram shows the hyperfine levels of the ground state ($^2S_{1/2}$) and the excited state ($^2P_{3/2}$) under a weak magnetic field. The 313 nm cooling and repumping lasers are indicated by blue and purple arrows, respectively. Microwave-driven state transfers are denoted by gray arrows. The red arrow marks the magnetically insensitive transition within the $^2S_{1/2}$ ground state.

SMA100B), routed through a microwave switch, amplified by a power amplifier, and directed to one of the central electrodes of the linear Paul trap. This electrode serves as an antenna, thereby coupling the microwave field to the trapped ions.

## III. EXPERIMENTAL PROCESS

First, $^9Be^+$ ions are produced via laser ablation, captured in the ion trap, and laser cooled to form a Coulomb crystal. Under a weak magnetic field, the hyperfine levels split into Zeeman sublevels, as illustrated in Fig. 2. Following the transition selection rules, a $\sigma^+$-polarized laser beam (blue in Fig. 2) propagating parallel to the magnetic field drives a closed cycling transition between the ground state $^2S_{1/2}$ $|F=2, m_F=2\rangle$ and the excited state $P_{3/2}$ $|F=3, m_F=3\rangle$, resulting in continuous photon scattering and a strong fluorescence signal. A repumping laser beam (purple in Fig. 2) pumps ions from the $F=1$ hyperfine sublevel of the ground state to the excited state $^2P_{3/2}$, from which they decay back to the $F=2$ hyperfine state via spontaneous emission. After laser cooling, the repumping laser is turned off 50 μs later than the cooling laser, resulting in the $^9Be^+$ ions being prepared in the $^2S_{1/2}$ $|F=2, m_F=2\rangle$ state and thus achieving the desired initial state for the experiment. Similarly, the state $^2S_{1/2}$ $|F=2, m_F=-2\rangle$ can be prepared using $\sigma^-$-polarized light.

To prepare $^9Be^+$ ions in the initial $|F=2, m_F=0\rangle$ state for exciting the magnetically insensitive transition, the population in the $|F=2, m_F=2\rangle$ state is first transferred to the $|F=1, m_F=1\rangle$ state and then to the $|F=2, m_F=0\rangle$ state. This is achieved using two resonant microwave $\pi$ pulses, as indicated by the solid gray arrows in Fig. 2. During the excitation of the magnetically insensitive transition, a fraction of the ions is excited to the upper state $|F=1, m_F=0\rangle$ and the ions remaining in the lower state $|F=2, m_F=0\rangle$ are transferred back to the $|F=2, m_F=2\rangle$ state via microwave transitions (dashed gray arrows in Fig. 2). Immediately afterward, the 313 nm cooling laser is turned on for fluorescence detection. The entire experimental sequence is illustrated in Fig. 3(a). A successful transition is identified by a reduction in fluorescence intensity, where the resonant condition corresponds to a distinct minimum in the fluorescence signal.

Prior to frequency scans of the magnetically insensitive transition, the resonance frequency and the optimized $\pi$-pulse duration of each microwave transition are individually determined by taking Rabi oscillation spectra, as illustrated in Fig. 3(b) for the transition $(F, m_F): |2,2\rangle \rightarrow |1,1\rangle$. By varying the microwave pulse duration at the resonant frequency, a Rabi oscillation spectrum as shown in Fig. 3(c) is obtained. Fitting these data yielded a coherence time of $\tau = 255(15)$ μs and a $\pi$-pulse duration of $t_\pi = 29.1(1)$ μs. All uncertainties reported in this paper represent standard uncertainties corresponding to one standard deviation ($1\sigma$). Compared with direct fluorescence detection, the transition detection technique integrating polarized cooling lasers and microwave state transfer effectively enhances the sensitivity of transition detection [30].

A typical microwave frequency scan of the magnetically insensitive transition is shown in Fig. 3(c), corresponding to a coil current of $I = 0.3$ A. The measured center frequency of the transition is 1250.031 04(12) MHz, with a full width at half maximum (FWHM) of 10.0(3) kHz. The displayed resonance profile represents a weighted average of 120 individual frequency scans. $^9Be^+$ ions, particularly those populated in the excited-state manifold during the laser cooling cycle, undergo chemical reactions with residual $H_2$ in the background gas to form $BeH^+$ [31], leading to a reduction in fluorescence signal due to ion loss. Therefore, prior to fitting, the data are corrected to account for the ion lifetime of $\tau = 248(20)$ s, which mitigates the spectral line-shape asymmetry caused by fluorescence decay. The ion lifetime is extracted from one-hour-long monitoring experiments following the timing sequence illustrated in Fig. 3(a).

The modest signal-to-noise ratio shown in Fig. 3(c) arises because the Zeeman splitting under weak magnetic fields is considerably smaller than the natural linewidth and Doppler broadening. Consequently, imperfect state preparation, state transfer, and excitation via magnetically insensitive transitions lead to unwanted population in the sublevels of the $F=2$ state, which in turn degrades the signal-to-noise ratio.

## IV. RESULTS AND DISCUSSION

Frequency scans of the magnetically insensitive transition have been conducted at different applied coil currents, with the measured resonance frequencies shown in Fig. 4. The current through the axial Helmholtz coils is varied from −0.2 A to 0.3 A in increments of 0.05 A, allowing multiple scans of the magnetically insensitive transition at varying magnetic-field strengths. The measurement order is randomized to reduce the influence of long-term magnetic-field drift. For each axial current setting, we compensated the residual radial field by scanning the currents of two radial coil pairs while monitoring the fluorescence of the closed cycling transition. The residual radial fields were minimized when the fluorescence intensity reached its maximum, corresponding to maximum fidelity. All experiments are carried out using Coulomb crystals of identical size and spatial configuration, thus minimizing systematic errors arising from spatial inhomogeneities in the magnetic field.

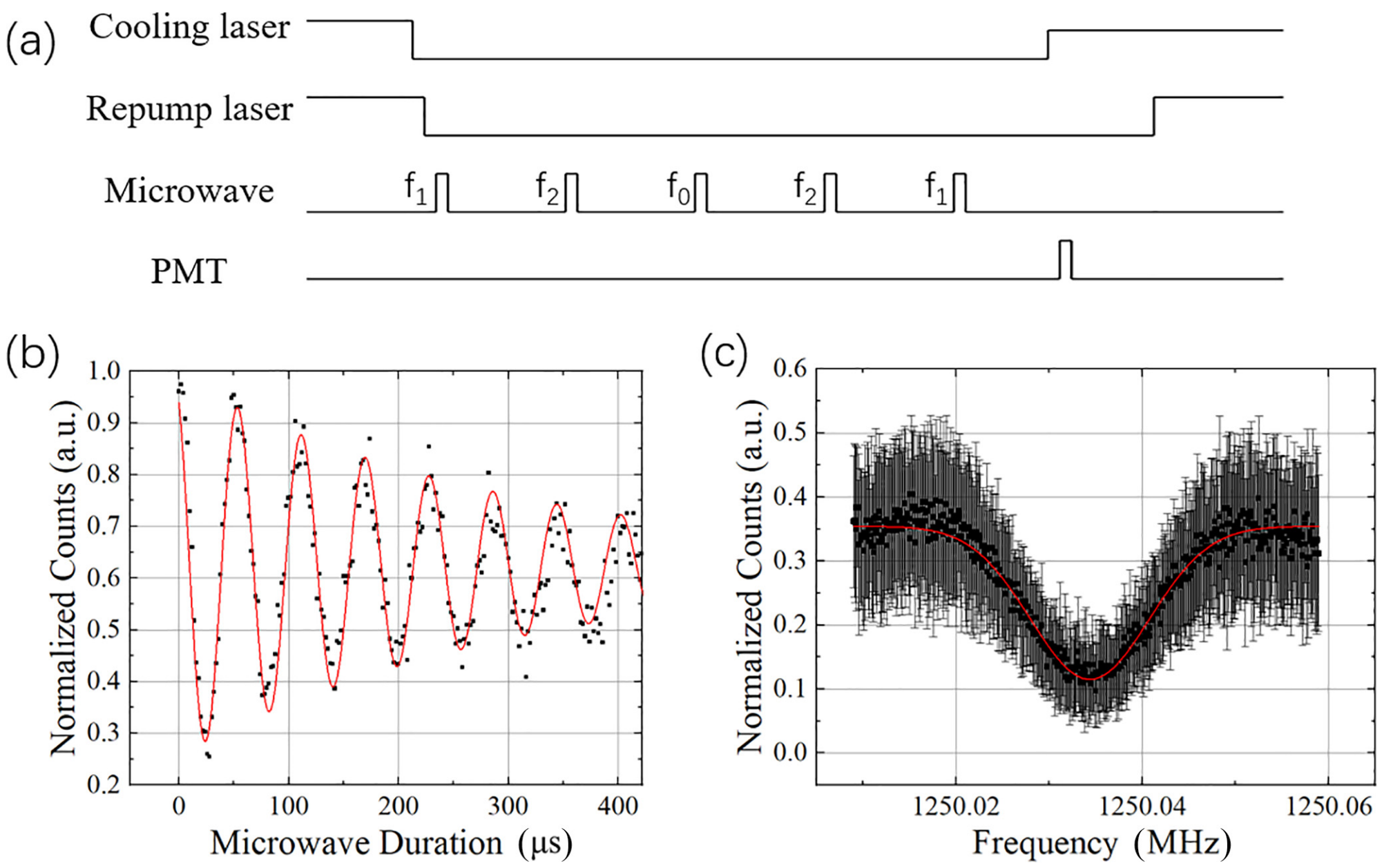

FIG. 3. (a) Timing sequence of a single laser-microwave multipulse experiment. The five microwave transitions correspond to frequencies $f_1$, $f_2$, $f_0$, $f_2$, and $f_1$, respectively, as indicated in Fig. 2. (b) Observed Rabi oscillation spectrum of the transition $(F, m_F) : |2,2\rangle \rightarrow |1,1\rangle$ at its resonant frequency of $f = 1235.042$ MHz, a coil current of $I = 1$ A, and microwave signal generator output power of $P = 237$ mV. (c) Observed microwave spectrum of the transition $(F, m_F) : |2,2\rangle \rightarrow |1,1\rangle$ at a coil current of $I = 0.3$ A. The spectrum is scanned over a 36 kHz range with a step size of 100 Hz. Each data point represents the average of 120 repeated measurements.

Under millitesla (mT) level magnetic-field strengths, the Breit-Rabi model can describe the hyperfine splitting with Zeeman splitting of the lithium-like ions, including the nuclear and quantum electrodynamic (QED) corrections. For the $|F = 2, m_F = 0\rangle \rightarrow |F = 1, m_F = 0\rangle$ transition, the transition frequency can be expressed as [13,24]

$$\nu_{00} = 2A\sqrt{1 + (g_J - g_I^{'})^2(1 + \delta)X^2}, \quad (1)$$

where $X = \frac{\mu_B B}{\Delta E_{\mathrm{HFS}}}$, $\Delta E_{\mathrm{HFS}} = 2hA$ for beryllium ions, $B$ denotes the magnetic-field strength, and $\mu_B$ is the Bohr magneton. The $g$ factors are defined by $g_J = -\mu_J/(J\mu_B)$ and $g_I{'} = -\mu_I/(I\mu_B)$, where $\mu_I$ and $\mu_J$ are the nuclear and electronic magnetic-moment operators, respectively. $\delta$ represents the correction to the $g$ factor, arising primarily from interelectronic-interaction effects [24]. Since $m_F = 0$, the contributions from the first-order Zeeman effect and all $m_F$-dependent corrections are eliminated [13,23,32].

To simplify the model, given that the magnetic-field strength is adjusted by the coil current, we define $k = (1 + \delta)[\frac{\mu_B(g_J - g_I{'})}{h} k_C]^2$ and the transition frequency can be written as

$$\nu_{00} = \sqrt{4A^2 + k(I - I_0)^2}, \quad (2)$$

where $k_C$ is the coil constant, defined as the magnetic-field strength produced per unit current at the position of the Coulomb crystal. $I_0$ denotes the coil current required for compensating the axial magnetic field, corresponding to a net zero magnetic field at the ion position.

By performing multiparameter fitting of $\nu_{00}$ (as a function of $I$) to Eq. (2), the hyperfine constant $A$ is determined to be $-625.008\,840(34)$ MHz, $k$ is 516.1(4.6) MHz$^2$/A$^2$, and $I_0$ is 0.0462(6) A. The fitting uncertainty is numerically equal to the standard error calculated by the principle of least squares. The goodness of fit, quantified by the coefficient of determination $R^2$, is 0.999 31. This $R^2$ value confirms the exceptional applicability of the Breit-Rabi model, which fully incorporates higher-order Zeeman effects, within our exper-imental regime. It should be noted that $k$ is not included as a fixed parameter input prior to fitting. In other words, our model remains valid when corrections to the Breit-Rabi model are considered and it is independent of the numerical values of $\mu_B$, $\delta$, $k_C$, and $g$ factors.

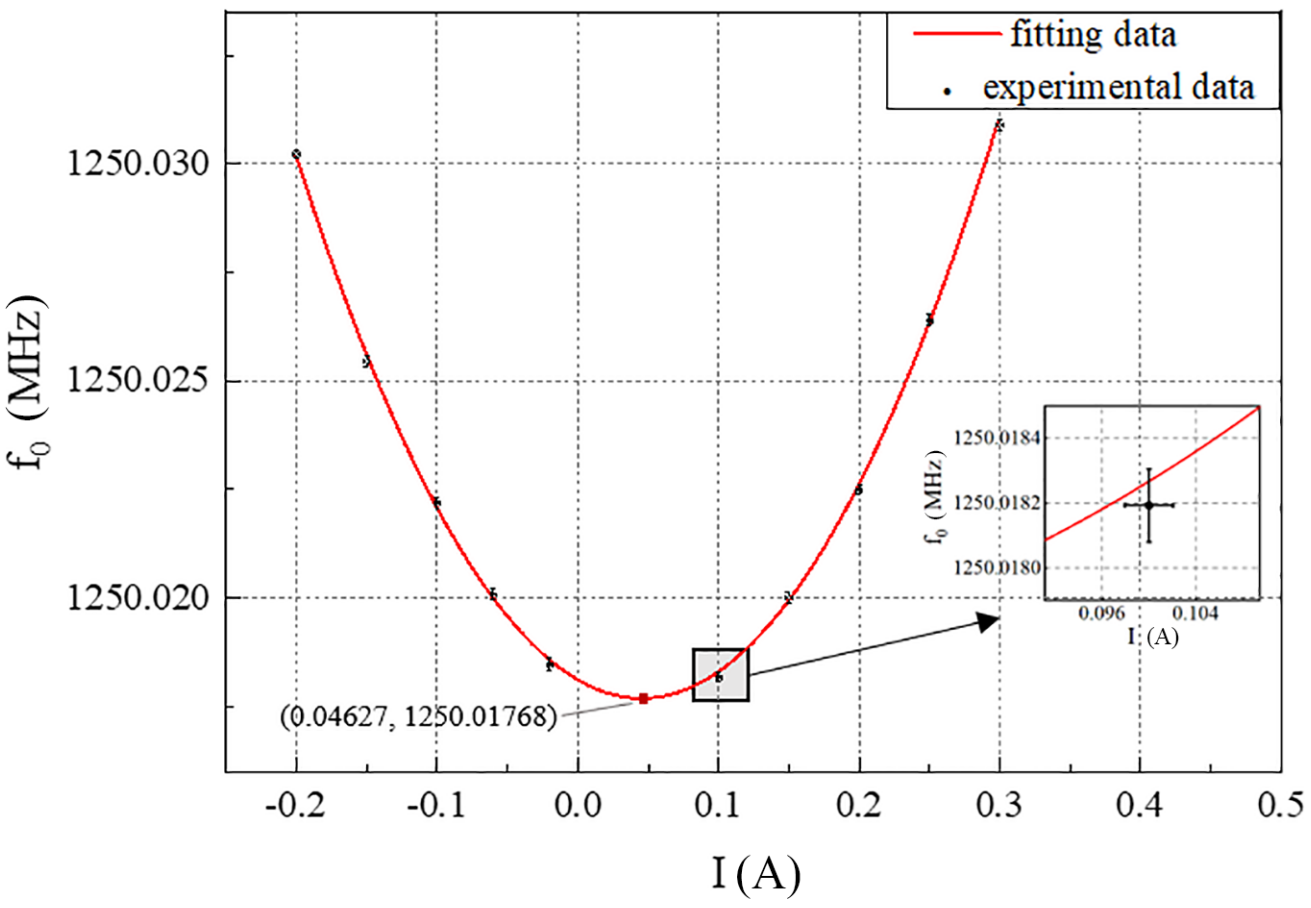

FIG. 4. Resonance frequencies of the $^2S_{1/2}$ $|F = 2, m_F = 0\rangle \rightarrow |F = 1, m_F = 0\rangle$ transition as a function of Helmholtz coil current. The error bars represent the uncertainties obtained from fitting the resonance curves. Current fluctuations are incorporated as horizontal error bars in the fitting process.

Our method for determining the hyperfine constant $A$ relies on fitting resonance frequencies measured at multiple near-zero magnetic-field strengths, thereby ensuring that errors from individual data points do not propagate directly into the determination of $A$. This approach effectively mitigates the impact of magnetic-field perturbations, including short-term fluctuations arising from output instability of the coil current source and long-term drift of the ambient magnetic field. The former are mitigated by averaging over 120 individual scans at each current setting; the latter are mitigated by randomizing the current settings. Resonance frequency uncertainty and electric current uncertainty are represented as vertical and horizontal error bars, respectively, in the data shown in Fig. 4. The overall magnetic field uncertainty from these sources is therefore incorporated into the statistical error budget of the fitting parameters.

The loss of $^9Be^+$ ions, primarily due to chemical reactions with background gas forming $BeH^+$ [31], leads to fluorescence decay and subsequent asymmetry in the spectral line shape. Prior to fitting the resonance peaks, the line shape is corrected to account for the $^9Be^+$ ion lifetime of 248(20) s. This correction results in a frequency shift of 17.4(1.4) Hz between the fitted resonance peaks before and after this correction. Thus the uncertainty in the ion lifetime contributes an uncertainty of 1.4 Hz into the determination of the central frequency.

Due to spatial magnetic-field inhomogeneity near the trap center and the finite size of the Coulomb crystal, the effective magnetic fields experienced by the individual ions within the Coulomb crystal are slightly different from one another. By adjusting the end-cap voltage to translate the Coulomb crystal position and measuring the magnetically sensitive transition $^2S_{1/2}\,|F=2, m_F=2\rangle \rightarrow |F=1, m_F=1\rangle$, we mapped the magnetic field near the trap center, yielding a magnetic-field gradient of $\sim$0.008 mT/mm. Combining this gradient with the crystal size of 210 μm, the magnetic-field inhomogeneity contributes an uncertainty of 0.02 Hz to the central frequency determination.

Since the microwave wavelength is far larger than the trapped ion's motional amplitude, the ions experience a phase-uniform driving field across their trajectory and are thus unaffected by the first-order Doppler effect. During microwave irradiation, the cooling laser is turned off. At an overestimate ion temperature of 100 K, the second-order Doppler shift arising from three dominant types of motion (secular motion, intrinsic micromotion driven by the RF field, and excess micromotion due to deviations of the ions from the RF nodal line) is calculated to be 0.058 Hz [33].

Frequency scans of the magnetically insensitive transition are performed at different microwave powers and ion trap RF power levels. However, at the present spectroscopic resolution of $\sim$120 Hz, no statistically significant dependence of the resonance frequency on either microwave or RF power is observed.

TABLE I. Estimation of systematic frequency shifts and uncertainties.

| Source of uncertainty | Shift (Hz) | Uncertainty (Hz) |
|---|---|---|
| Statistical error | 0 | 34 |
| Magnetic-field inhomogeneity | 0 | 0.02 |
| Second-order Doppler | 0.058 | 0 |
| Microwave signal generator (MSG) | 0 | 0.003 |
| Lifetime | 17.4[a] | 1.4 |
| Total | 0 | 35 |

[a]Prior to fitting resonance peaks, the spectral line shape was corrected to account for the $^9Be^+$ ion lifetime.

The contributions to the uncertainty budget for the hyperfine constant $A$ are summarized in Table I. Considering all the systematic shifts and uncertainties, the final value of the hyperfine constant $A$ is determined to be $-625.008\,840(35)$ MHz, with a total uncertainty of 35 Hz, corresponding to a relative precision of $5.6 \times 10^{-8}$.

A comparison of the experimental results with those from other research groups and theoretical predictions is presented in Table II. For clarity, only shifts and uncertainties exceeding $10^{-3}$ Hz are listed in the Table I. For example, the shift due to residual radial fields is estimated to be less than 1 nHz and the AC Zeeman shift from the RF trap drive, which is less than 10.5 μHz, is not included. The currently most precise value [13] is derived from measurements in strong magnetic fields on the order of several tesla. In this work, the ground-state hyperfine constant $A$ of $^9Be^+$ is directly determined via measurements of the magnetically insensitive transition under magnetic-field strengths within a range of $\pm$0.5 mT centered at zero magnetic field, achieving an uncertainty of 35 Hz—representing a one-order-of-magnitude improvement in precision over the previous weak-field study [27]. Our result is consistent with those from strong-field measurements within respective uncertainty budgets.

The current theoretical value for the ground-state hyperfine constant $A$ of $^9Be^+$ ion is calculated under the point-nucleus approximation [1,10]. The deviation between our experimental measurement and these theoretical values can be attributed to the finite nuclear size effect, from which the effective nuclear Zemach radius can be extracted [34]. This yields an effective nuclear Zemach radius of 4.03(5) fm, with its precision constrained by the accuracy of theoretical calculations.

## V. CONCLUSIONS

A high-precision, high-contrast laser-microwave spectroscopy system has been developed to measure the ground-state hyperfine constant of $^9Be^+$. By precisely adjusting the current of the Helmholtz coils, we perform precise frequency measurements of the magnetically insensitive transition $^2S_{1/2}$ $|F=2, m_F=0\rangle \rightarrow |F=1, m_F=0\rangle$ of $^9Be^+$ ions over a magnetic-field range of $\pm$0.5 mT centered at zero magnetic field. The ground-state hyperfine constant $A$ is derived by fitting the experimental data to the Breit-Rabi model, yielding $A = -625.008\,840(35)$ MHz with a relative precision of $5.6 \times 10^{-8}$. Fitting resonance frequencies measured across multiple magnetic-field strengths effectively suppresses the dominant error contribution from magnetic-field uncertainties. Furthermore, the weak magnetic-field conditions also strongly suppress corrections deviating from the Breit-Rabi model. This result represents the most precise measurement of $A$ for $^{9}Be^{+}$ under weak magnetic fields reported to date.

TABLE II. Comparison of experimental and theoretical values of the ground-state hyperfine constant $A$ for $^{9}Be^{+}$.

| Research group and year | Magnetic field | $A_{S1/2}$ (MHz) |
|---|---|---|
| D. J. Wineland *et al.* (1983) [12] | 1.134 T | −625.008 837 048(10) |
| K. Okada *et al.* (1998) [27] | 0.7 mT | −625.008 82(26) |
| T. Nakamura *et al.* (2002) [26] | 0.47 T | −625.008 835 23(75) |
| J. J. Bollinger *et al.* (2011) [13] | 4.4609 T | −625.008 837 044(12) |
| V. A. Yerokhin (2008) [35] | Theoretical[a] | −625.08(2)<br>−625.11(3) |
| K. Pachucki *et al.* (2014) [10] | Theoretical | −625.3927(36)(16) |
| Our work | 0–0.5 mT | −625.008 840(35) |

[a]Two distinct $A$ values were obtained by incorporating different nuclear magnetic moment models.

Combining the point-nucleus approximation with our experimental results, we determine the effective nuclear Zemach radius to be 4.03(5) fm. Future improvements in theoretical calculations will further enhance the precision of this value.

Notably, in Table II, a 2.4$\sigma$ discrepancy exists between the values reported by Bollinger *et al.* [13] and Nakamura *et al.* [26], potentially indicating either unresolved corrections associated with magnetic fields or unresolved systematic errors under high magnetic-field conditions. Future theoretical and experimental investigations will help pinpoint the underlying causes of this discrepancy.

Further improvements in experimental precision can be achieved by enhancing magnetic-field control, including both passive shielding and active compensation. Employing a multilayer enclosure fabricated from high-permeability materials can significantly attenuate external magnetic fields. Real-time magnetic-field compensation can be implemented by monitoring the field with a high-precision magnetometer and using a feedback control system to adjust the current in compensation coils, thus canceling out dynamic fluctuations.

In addition, $^{9}Be^{+}$ ions are widely used as coolant species in sympathetic cooling experiments and they can also serve as *in situ* sensors [36]. By utilizing the ground-state hyperfine transition of $Be^{+}$ ions as a high-precision magnetic-field probe, the magnetic-field strength at the trap center can be precisely calibrated in real time, thus enabling the quantification of the Zeeman shift induced by this field experienced by the target spectroscopic ions.

## ACKNOWLEDGMENTS

This work was supported by the National Key R&D Program of China (Grant No. 2021YFA1402103) and the National Natural Science Foundation of China (Grant No. 12393825).

## DATA AVAILABILITY

The data that support the findings of this article are not publicly available. The data are available from the authors upon reasonable request.

*Correction:* A minor error in Eq. (2) and a value in the text following Eq. (2) have been fixed.